 	\documentstyle[12pt]{article}

	\newcommand{\mlc}{\multicolumn}
	\newcommand{\beq}{\begin{equation}}
	\newcommand{\eeq}{\end{equation}}
	\newcommand{\ber}{\begin{eqnarray}}
	\newcommand{\eer}{\end{eqnarray}}

 	\newcommand{\lag}{\langle}
 	\newcommand{\rag}{\rangle}
 	\newcommand{\prn}{\sc prn}
 	\newcommand{\logma}{\sc logmap}

\begin{document}
	\begin{titlepage}
	\begin{center}

	\hfill IP/BBSR/93-52

	\vskip .5in

	{{\Large \bf Logistic Map: A Possible Random Number Generator}}

	\vskip .15in

	S. C. Phatak\footnote{\sl phatak@iopb.ernet.in} and
	S. Suresh Rao\footnote{\sl suresha@iopb.ernet.in}\\[.15in]
	{\em Institute of Physics, Bhubaneswar-751005, India\\}
	\end{center}

	\vskip .15in

	\begin{abstract}
\noindent The logistic map is one of the simple systems
exhibiting order to chaos transition. In this work we have
investigated the possibility of using the logistic map in the chaotic
regime ({\logma}) for a pseudo random number generator. To this
end we have performed certain statistical tests on the series of
numbers obtained from the {\logma}. We find that the {\logma} passes these
tests satisfactorily and therefore it possesses many properties
required of a pseudo random number generator.
\vskip 1 cm

\noindent PACS number(s): 05.45 +{\bf b}, 05.20, 01.50.H, 02.70Lq
	\end{abstract}
	\end{titlepage}

	\newpage
	\renewcommand{\thepage}{\arabic{page}}
	\setcounter{page}{1}

	\section {Introduction}
	\setcounter{equation}{0}

	\baselineskip 18pt

A sequence of numbers that are chosen at random are useful in many
different kinds of applications like simulation, sampling,
numerical analysis, decision making, recreation etc. A sequence of
truly random numbers is unpredictable and hence irreproducible.
Such a sequence can only be generated by a physical process, for
example, radioactive decay, thermal noise in electronic devices,
cosmic ray arrival time etc. In practice, however, it is very
difficult to construct physical generators which are fast enough
and at the same time accurate and unbiased. Furthermore, one
would like to repeat the calculation at will, for debugging or
developing the program. Thus for most calculational purposes,
pseudo random numbers ({\prn}) have been introduced. Pseudo random
numbers are numbers computed from a deterministic
algorithm (hence it is called pseudo random or quasi-random) and
therefore reproducible. Obviously these are not at all random in the
mathematical sense, but are supposed to be indistinguishable from
a sequence generated truly randomly. A good {\prn} generator should
possess long period, high speed and randomness.

Over the years various {\prn} generators have been developed and
can be broadly classified into the following categories~\cite{james}:
those are, 
i.) Linear recurrence methods,
ii.) Multiplicative congruential generators,
iii.) Tausworthe generators and
iv.) Combination generators.
All these are bit based generators. These methods have various
parameters or inputs and the period and the statistical
properties of {\prn} sequences sensitively depend on these parameters.
The first two are known to
have periods of about \(\sim 10^9\) on a \(32-\)bit machine. In
the other two methods one can achieve much larger period (\(\sim
10^{170}\))~\cite{james}.

It is not easy to invent a foolproof source of random numbers.
Generally a number of tests~\cite{knuth} are performed to test
the {\em randomness} of the numbers generated by {\prn}'s. In spite
of these tests, one finds that a {\prn} that has passed these tests
may fail when applied to some physical applications: For
example, in a recent study Ferrenberg ~{\it et al.}~\cite{feren}
have shown that even the high quality {\prn}s are
biased under certain circumstances. Extensive Monte Carlo
simulations by this group on an Ising
model, for which exact answers are known, have shown that
ostensibly high quality random number generators may lead to
subtle, but dramatic, systematic errors for some algorithms, but
not others. They traced the discrepancy to the correlations in
the random numbers. Another recent study by Vattulainen {\it et
al.}~\cite{nissi} find no such correlations. This only means
that what is ``random" enough for one application may not be
{\em random} enough for another. The important criterion is that a
specific algorithm must be tested together with the random
number generator being used regardless of the tests which the
generator has passed.
This being the scenario, it may be useful to consider {\prn}'s
based on algorithms different from conventional algorithms.

In the present work we discuss a {\prn} generator based on an inherently
chaotic (random) algorithm and its statistical properties.
Here we have employed the logistic map in the chaotic regime
as a {\prn} generator. We feel that such an effort is useful
because it provides an
entirely different method of producing {\prn}'s. Also the algorithm
used is very simple, so the generator is quite fast.
There have been earlier attempts~\cite{prak} to use the logistic map
as a random number generator. For example, the
logistic map in the chaotic regime was used to calculate the
properties of a well-known random process; the invasion percolation
problem. In this study, it was found that the static properties of
percolating clusters, except for the percolation threshold, are
correctly calculated but the dynamical properties are not. It is not
clear from this work if this difference is due to non-uniform nature
of the distribution or due to the correlation between successive
numbers. The method we are adopting here differs from the
earlier calculations in two respects. In the earlier works the random
numbers were drawn from the logistic map which is known to have a
distribution of \(\sim \sqrt{x (1-x)}\) and
used in the simulation studies. We use a simple transformation to
convert this distribution into a uniform distribution. Also, the
earlier calculations make no attempt to either
remove or study the effects of the correlations that exist between
succesive numbers generated by the logistic map. We have done that.
Here we draw the numbers from a
uniform distribution (see below) and investigate the statistical properties.
Our calculations show that the period of such a generator is
certainly larger than \(10^9\) (although theoritically infinite) if the
computations are done in double precision. Our investigation of
the distribution properties of this {\prn} generator shows some
peculiarities. These can be cured by introducing the
\(\tau\)--shift (to be explained below).
In the following we shall first describe the logistic map and
discuss the {\prn} generator in Section 2. The results of different
tests performed on {\prn}'s thus generated are presented in Section
3 and the conclusions are given in Section~ 4.
	\section{The Logistic Map}
	\setcounter{equation}{0}
Chaos in dynamical systems has been investigated over a
long period of time~\cite{gener}. With the advent of fast
computers the numerical investigations on chaos have increased
considerably over the last two decades and by now, a lot is
known about chaotic systems. One of the simplest and most
transparent system exhibiting order to chaos transition is
the logistic map~\cite{feig}. The logistic map is a
discrete dynamical system defined by~\cite{feig},
\beq
 x_{i+1} = \mu x_i (1 - x_i ) \label{eq:logm}
\eeq
with \(0 \leq x_i \leq 1\). Thus, given an initial value (seed)
\(x_0\), the
series \(x_i\) is computed. Here the subscript \(i\) plays the role of
discrete time. The behavior of the series as a function of the
parameter \(\mu\) is interesting. A thorough investigation of
logistic map has already been done[7]. Here, without going into
detailed discussion, we simply note that,
\begin{itemize}
\item Eq.(\ref{eq:logm}) has \(x=0\) and \(x=(\mu-1)/\mu\) as fixed points.
That is,
if \(x_i =0\) or \((\mu-1)/\mu\), then \(x_{i+1} =x_i\).
\item For \(\mu < 1\), \(x=0\) is an attractive (stable) fixed point.
That is, for any value of the seed \(x_0\) between \( 0\) and
\(1\), \(x_i\) approaches \(0\) exponentially.
\item For \(1\leq \mu \leq 3\), \(x=(\mu-1)/\mu\) is an
attractive fixed point.
\item For \(3< \mu < 4\), the logistic map shows interesting
behavior such as repeated period doubling, appearance of odd
periods etc.
\item For \(\mu = 4\) the logistic map is chaotic.
\end{itemize}
Since the chaotic behavior of the logistic map is of
interest to us, we shall discuss the last point in detail.
If we choose two
seeds \(x_0\) and \(y_0 =x_0 +\delta x\), with \(\delta x\)
arbitrarily small, \(x_i\) and \(y_i\) differ by finite amount
for large enough \(i\). For example, if
\(\delta x =10^{-12}\), \(x_i -y_i \sim 0.1\), for \(i\sim 40\).
In fact, \(x_i -y_i\) grows exponentially with \(i\) and this is
the definition of chaos. An analytic solution of logistic map
exists for \(\mu=4\). If we choose
\(x_i =(1-\cos(\theta_i) )/2\) and \(x_{i+1}
=(1-\cos(2\theta_{i}))/2\) the logistic map for \(\mu=4 \)
can then be defined by
\beq
\theta_{i+1}= \left \{ \begin{array}{ll}
 2 \theta_i, & \mbox {for $\theta_i < \pi/2$}\\
 2 \pi - 2 \theta_i, & \mbox{for $ \theta_i > \pi/2$.}
\end{array}\right. \label{eq:pifold}
\eeq
Clearly, the map in terms of \(\theta_i\) is given by stretching
the line of length \(\pi\) to \(2\pi\) and folding it. An
examination Eq.(\ref{eq:pifold}) shows that one gets periodic series if the
seed \(\theta_0\) is a rational fraction of \(\pi\). On the
other hand, the series does not have periodicity if \(\theta_0\)
is an irrational fraction. Also, for any \(\theta_0\) which is an
irrational fraction of \(\pi\), the set \(\{\theta_i\}\)
computed using Eq.(\ref{eq:pifold}) is distributed uniformly
between \(0\) and
\(\pi\). It is this property of the logistic map that we intend
to exploit for generating uniformly distributed random numbers
and to study its statistical behavior.

Consider a set of numbers
\beq
y_i \> = \> \frac{\cos^{-1}{(1-2x_i)}}{\pi} \label{eq:ya}
\eeq
where,
\beq
x_{i+1} \> = \> 4 x_i (1-x_i). \label{eq:yb}
\eeq
 From the discussion above, \(y_i\)'s are expected to be
distributed uniformly between \(0\) and \(1\), except when
\(x_i \ = \ \frac{1}{4}, \ \frac{1}{2}, \ \frac{3}{4} \ \mbox{or }
\ 1\). For these special values one gets periodic series. For any
other rational fraction \(x_0\) we do not expect any
periodicity. Thus it appears that, starting with a rational
fraction $x_0$ we can generate a set of uniformly distributed
numbers $\{y_i\}$ using Eqs(\ref{eq:ya},\ref{eq:yb}).
Furthermore since a small
change in $x_0$ produces large deviations in $x_i$'s (and
therefore $y_i$'s) different initial values of $x_0$ differing
by small amount would produce different uncorrelated sets $\{y_i\}$.
This seems to be a very good
property for a random number generator to possess.

There are however two difficulties to be overcome before
we can construct a random number generator from Eqs(\ref{eq:ya},
\ref{eq:yb}). The
first is due to truncation error in computers. We find that,
when the calculations are done in single precision, $x_i$
becomes $0$ after about $5000$ iterations. The exact value of
$i$ depends on $x_0$ but it is ususally $\sim 5000$. The reason
for this is as follows. If $x_i$ differs from $1/2$ by a small amount
$\epsilon$ ($x_i=0.5+\epsilon$), $x_{i+1}=1-\epsilon^2$ and if
$\epsilon^2 < 10^{-7}$, $x_{i+1}$ is stored as $1$ in the
computer. So $x_{i+n}=0$ for $n\geq2$. This difficulty is
overcome by either doing the calculation in double precision (higher
precision, if possible) or by modifying the algorithm suitably, when
\(x_i\) is close to \(0.5\). We find that this problem does not arise for upto
\(\sim 10^9\) iterations when the calculation is done in double precision.

The second difficulty is that of correlation among the
successive \(y_i\)'s. Eq(2.2) clearly shows that the succesive
$y_i$'s are correlated, although a set of $y_i$'s are uniformly
distributed in the interval [0,1]. The standard procedure of removing
such a correlation is to shuffle the set of numbers obtained from
logistic map. Our calculations show that the correlations have
peculiar effect on the distribution properties and these persist even
after shuffling (see later). On the other hand, the correlations
between two numbers $y_i$ and $y_{i+\tau}$ reduce as $\tau$ is increased
from 1. We call these as $\tau$-shifted numbers. The statistical
tests have been performed on these $\tau$-shifted set with $\tau$
ranging from 1 to 14. The following calculations have been done in double
precision.

 \section{Distribution Tests}
	\setcounter{equation}{0}

In order to test the {\prn} generator ({\logma}) described in Section 2,
we have performed certain tests, for various values of \(\tau\),
described below. For comparison we
have also done these tests on {\prn} generator RAND-- available on our
machine and RANMAR-- the generator proposed by Marsaglia and Zaman.
These tests are:

\begin{enumerate}
\item Distribution test: here we verify Central Limit Theorem.
\item Moments Caclulation: We have calculated \(\lag x^n\rag\)
and their variance \(\sigma_n\).
\item \(\chi^2\)-test: In the 1--d case, we have divided the
interval \([0,1]\)
into $n$ equal bins and calculated $\chi^2$ and its
distribution. For the 2--d case, we have divided the region
\([0,1] \otimes [0,1] \) into \(n\) equal blocks and repeated
the calculation of \(\chi^2\) and its distribution.
\end{enumerate}
In addition, we have also performed other tests~\cite{knuth}
such as run-up test, $n$-tuple test etc. Results of these tests
will not be presented here. The main reason is that the results
of these tests are in concurrence with the above mentioned tests
and it will not alter our conclusions. In the following, we report
the results for {\logma} and RANMAR.

\subsection{Distribution Test}

Suppose that \(\{x_i \}\) is a sequence of mutually independent
random variables that are governed by the probability density function
\(P(x)\). Then the Central Limit Theorem asserts that, subject to
certain conditions on the moments of \(P(x)\)~\cite{kinch} the variable
\(y_N=\sum_{i=1}^{N} x_i\), in the limit of large \(N\), are distributed
normally: i.e.,
\[
P_{N}(y)=\frac{1}{\sqrt{2 \pi \sigma^2}} \exp{-\frac{(y-\mu)^2}{2
\sigma^2}}, \]
\[
{\rm where}\quad \mu = N \lag x \rag \quad {\rm and}\quad
\sigma^2 = N (\lag x^2 \rag - {\lag x \rag}^2).
\]

We have obtained the distribution function of the numbers obtained
from {\logma} for \(N = 24\) for various \(\tau\)--shifts.
In fig.(1), the soild line indicates the
exact distribution and diamond dots indicate the distribution
obtained from the {\logma} with \(\tau = 1, 2 \hbox { and }\> 3\)
calculation. From fig.(1), one can see that for \(\tau = 1\)
(sequential case),
the distribution from the {\logma} is skewed a little. Though the
distribution for \(\tau = 2\) is better agreed than the
sequential case, still it is far from satisfactory. For
\(\tau = 3\) and greater, the distribution is in excellent
agreement with the theoretical one. Thus this test shows that,
although the successive numbers are correlated, a set of numbers
obtained by picking every alternate or every third number, would
behave as a set of random numbers~\cite{note}.

\subsection{Moments}

For a given set of \(\tau\)--shifted $N$ numbers generated
from {\logma}, we
calculate $n^{\rm th}$ moment $\lag x^{n}_{\tau}\rag$ as
\beq
\lag x^{n}_{\tau} \rag = \frac{1}{N} \sum_{i=1}^{N} x_{i+\tau}^n.
\eeq
We then repeat this test for $M$ such sets and caclulate the global
average $\overline {\lag x^{n}_{\tau} \rag}$ and variance $\sigma_n$ as
\ber
\overline {\lag x^{n}_{\tau} \rag} &=& \frac{1}{M} \sum_{j=1}^{M} {\lag
x^{n}_{\tau} \rag}_j \\
\sigma_n(\tau) & = & \frac{1}{M} \sum_{j=1}^{M} {\lag
x^{n}_{\tau} \rag}_{j}^2- {\overline {\lag x^{n}_{\tau} \rag}}^2
\eer
We have varied \(\tau\) from 1 to 14 for two different values of
$N$--1024 and 8096 and $M$ is held fixed at
$4000$. For random numbers uniformly distributed between $0$ and
$1$ ${\overline {\lag x^n \rag}}^{\rm th}=\frac{1}{n+1}$ and
$\sigma_{n}^{\rm th}=\frac{n^2}{(2n+1){(n+1)}^2}$. The results for the
{\prn} generator based on logistic map are shown in Table 1.

An inspection of Table 1 shows that $\overline {\lag x^{n}_{\tau}
\rag}$ agrees well with the theoretical values for different n's and
$\tau$'s. The agreement is comparable with the results of RANMAR. But
$\sigma_n(\tau)$ departs significantly from the theoretical value for
$\tau = 1$, and the departure systematically increases with the
increase in n. The variance is calculated by subtracting two almost
equal numbers, so some loss of accuracy is expected. But the
departure is much larger. On the other hand, for $\tau$ larger than
4, the calculated variances are close to the theoretical values and
are comparable with those obtained from RANMAR.

\subsection{$\chi^2$-Test}

In 1--D $\chi^2$ test, the interval [0,1] is divided into n equal
parts and the numbers $r_i$ falling in $i^{th}$ interval, out of a
set of N numbers is calculated. For a uniform distribution, $\chi^2$
is defined by,
\beq
\chi^{2}_n=\sum_{i=1}^{n}\frac{(r_i-N/n)^2}{N/n}
\eeq
The test is repeated for $M$ such sets
and the distribution of $\chi_n^2$ is obtained. Theoretically,
the $\chi^2$ thus obtained should have a $\chi^2$
distribution for $n-1$ degrees of freedom:
\beq
P_{\rm th}(\chi^2_n) \>= \> \frac{(\chi^2_n)^{\frac{n-3}{2}}
e^{-\chi^2_n}}{\int d\chi^2_n (\chi^2_n)^{\frac{n-3}{2}}
e^{-\chi^2_n}}.
\eeq
The calculation is done for $N=10240$, $M=4000$, $n$ is
varied from $2$ to $256$ and $\tau$ is varied from 1 to 14. The
results for $n = 4, 64\hbox { and }256$ are displayed
in figs(2a,2b and 2c). The corresponding \(\chi^2_n\) for RANMAR is also
given for comparison.

Consider $\tau = 1$ or the sequential case first. The calculated
$\chi^2$ distribution
differs significantly from the theoretical one. This is somewhat
surprising, since the set of numbers obtained from the logistic map
are are uniformly distributed. We have confirmed that shuffling does
not mitigate this problem. Thus the correlations between successive
numbers are probably responsible for this behavior.

As $\tau$ is increased, the agreement between calculated and
theoretical distributions improves. From fig.(2a), we find that for
$n = 4$, $\tau = 2$ distribution already agrees reasonably with the
theoretical distribution. On the other hand, for $n = 256$, one gets
a good agreement for $\tau \sim 6$ or larger. This clearly shows
that, in order to have `randomness' on a finer scale (corresponding
to smaller bin size, or larger n), the $\tau$-shift must be larger.

For 2--D $\chi^2$--test, we choose a pair of succesive numbers (x and
y coordinates) and determine how they are distributed in $n$ equal-area
blocks covering a square of unit side. The $\chi^2$ distributions
are calculated as discussed in the 1--D case and the results for
$n = 4, 64 {\hbox{ and }}256$ are presented in figs(3a, 3b and 3c).
These results
follow the same pattern as that of 1--D case with one difference. For
$\tau = 1$, the calculated $\chi^2$ distribution is nowhere near the
theoretical distribution for \(n > 4\). This is simply because the succesive
numbers are highly correlated.
This can be explained as follows. For the sequential case, when the unit
square is divided into four equal blocks, each block would have some number of
points. Hence for \(n = 4\), one does not expect any abnormal behavior in
the \(\chi^2\) distribution. However, this is not the situation when the
unit area is divided into more than \(2\times2\) blocks; it would so happen
that some of the blocks do not contain any points at all. The contribution of
these blocks to the \(\chi^2\) (see eq.(3.4)) would be \(N/n\), thus
pushing the value of \(\chi^2\) higher.
However, with a $\tau$--shift of 4 or
larger, these correlations are more or less wiped out and one gets
reasonably good agreement between calculated and theoretical
distributions.

\section{Summary and Conclusions}

Various tests on the series of numbers obtained from the {\logma}
have been performed in this work. These tests bring
out certain peculiarities which have not been noted before. We notice
that,
\begin{enumerate}
\item the distribution test does not satisfy the central limit theorem
for \(\tau=1\). The correlations between succesive numbers, which are known to
exist, are probably responsible for this. But for \(\tau>2\), the agreement
with the central limit theorem is excellent.
\item The moments calculation comfirms the result of the distribution
test. That is the calculated moments agree with their theoretical
values, even for $\tau=1$. The variances of
these moments disagree with their theoretical values for $\tau = 1$.
However, for $\tau > 4$, the moments as well as variances agree with
the theoretical values. This shows that the correlations between
succesive numbers play a subtle role in moments calculation and
removal of these correlations (by $\tau$--shifting) is essential.
\item the $\chi^2$--tests show that, for $\tau = 1$ the
$\chi^2$--distributions systematically differ from the theoretical
distribution. The agreement between calculated and theoretical
distributions is improved by increasing $\tau$. Thus, one must use
$\tau$--shifted numbers (with larger value of $\tau$ for smaller
interval-size) to obtain acceptable $\chi^2$--distribution.
\end{enumerate}

As far, computer time, the {\logma} is about three times slower than the RANMAR
However, in the calculations where {\prn}'s are
used, a relatively small fraction of time is spent in generating random
numbers. Therefore the relative slowness of the {\logma} is not a big
handicap. Furthermore, independent sets of random {\prn}'s can be easily
obtained from the {\logma} by choosing different values of seeds \(x_0\)'s.
This is so because two seeds differing by a small amount would generate
entirely different series after a few iterations. This property makes the
{\logma} easily adaptable to parallel processing.

To conclude, we note that $\tau$--shifted {\logma}
satisfies some of the elementary tests a pseudorandom number
generator must pass. The {\logma} being based on a physically chaotic
process, the calculations where randomness, as opposed to computer time is
important, it is advantageous to use {\logma}.

{\Large {\bf {Acknowledgements}}}

\noindent We gratefully acknowledge many useful discussions with
Profs. R. E. Amritkar, S. M. Bhattacharjee, A. M. Jayannavar and A. Khare.

\newpage
{\Large {\bf Figure Captions}}

fig.(1) Central Limit Theorem Verification for $24$ numbers. The solid
curve is obtained from the exact distribution
\(P_{N}(y)=\frac{1}{\sqrt{2 \pi N \sigma^2}} \exp{-\frac{(y-N\mu)^2}{2
N\sigma^2}},\hbox{ where }N = 24,\\ \mu = 1/2 \hbox{ and }
\sigma^2= 1/12\). The diamonds with dot are obtained from the {\logma}
with \(\tau = 1,2 \hbox{ and } 3\) (as indicated by the label).

fig.(2) The \(\chi^2\) distribution of the {\logma} for the $1-$dimension
case. The smooth curve is obtained from the exact distribution, eqn.(3.5)
and the histogram is obtained from the {\logma} and the RANMAR, labelled
appropriately. a) for $4$ bins, b) for $64$ bins and c) for $256$ bins.

fig.(3) The \(\chi^2\) distribution of the {\logma} for the $2-$dimension
case. The smooth curve is obtained from the exact distribution, eqn.(3.5)
and the histogram is obtained from the {\logma} and the RANMAR, labelled
appropriately. a) for $4$ bins, b) for $64$ bins and c) for $256$ bins.
\newpage
\begin{table}
\begin{tabular}{||l|c|c|c|c|c|r||} \hline \hline
 \mlc{2}{||c|}{$n$} & \(\tau\>=\>1\)& \(\tau\>=\>3\)& \(\tau\>=\>6\)
& RANMAR & exact \\ \hline
 1&$ \lag x \rag$ & .50010& .50006& .50000& .49993& .50000 \\
  &$ \sigma_1   $ & .08400& .08460& .08626& .08155& .08333 \\
\hline
 2&$ \lag x^1 \rag$ & .33340& .33341& .33332& .33324& .33333 \\
  &$ \sigma_2     $ & .03736& .08784& .09102& .08736& .08889 \\
\hline
 3&$ \lag x^3 \rag$ & .25004& .25007& .24999& .24991& .25000 \\
  &$ \sigma_3     $ & .01392& .07828& .08111& .07964& .08036 \\
\hline
 4&$ \lag x^4 \rag$ & .20002& .20006& .20000& .19992& .20000 \\
  &$ \sigma_4     $ & .00518& .06839& .07092& .07086& .07111 \\
\hline
 5&$ \lag x^5 \rag$ & .16667& .16672& .16667& .16660& .16667 \\
  &$ \sigma_5     $ & .00273& .06002& .06240& .06305& .06313 \\
\hline
 6&$ \lag x^6 \rag$ & .14285& .14291& .14286& .14280& .14286 \\
  &$ \sigma_6     $ & .00276& .05318& .05550& .05642& .05651 \\
\hline
 7&$ \lag x^7 \rag$ & .12498& .12505& .12500& .12495& .12500 \\
  &$ \sigma_7     $ & .00363& .04759& .04990& .05087& .05104 \\
\hline
 8&$ \lag x^8 \rag$ & .11109& .11115& .11112& .11107& .11111 \\
  &$ \sigma_8     $ & .00469& .04298& .04528& .04620& .04648 \\
\hline
 9&$ \lag x^9 \rag$ & .09997& .10004& .10001& .09997& .10000 \\
  &$ \sigma_9     $ & .00567& .03913& .04143& .04225& .04263 \\
\hline \hline
\end{tabular}
\caption{$\lag x^n\rag$ and $\sigma_n$ for {\logma} with $\tau = 1, 3
\hbox{ and }6$ and RANMAR for\hfill\break
$n=1,2,\cdots,10$. $N=8196$ and $M=4000$.}
\end{table}

\begin{thebibliography}{99}
\bibitem{note} We find that for \(\tau=1\) and with shuffling, the
agreement with the theoretical distribution is very good. This indicates
that the departure for \(\tau=1\) case is due to the correlation between
successive numbers. However, the shuffling does not improve the results for
moments and \(\chi^2\) tests.
\end{thebibliography}
\end{document}